\documentstyle[emulateapj,apjfonts,psfig]{article}

\newcommand\cxo{{\em Chandra}}
\def\tornado{G357.7--0.1}

\def\Msun{\ifmmode{M_\odot}\else{$M_\odot$}\fi}
\newcommand\HI{H\,{\sc i}}
\newcommand\HII{H\,{\sc ii}}

\newcommand\etal{{\rm et~al.\ }}

\lefthead{GAENSLER ET AL}
\righthead{\uppercase{X-RAY EMISSION FROM THE TORNADO}}

\begin{document}
\title{Untwisting the Tornado: X-ray Imaging and Spectroscopy of G357.7--0.1}
\submitted{Accepted to ApJ Letters on 2003 Jul 08}
\author{B. M. Gaensler,\altaffilmark{1} J. K. J. Fogel,\altaffilmark{2}
P. O. Slane,\altaffilmark{1} J. M. Miller,\altaffilmark{1} \\
R. Wijnands,\altaffilmark{3} 
S. S. Eikenberry,\altaffilmark{4} and
W. H. G. Lewin\altaffilmark{5}}
\altaffiltext{1}{Harvard-Smithsonian
Center for Astrophysics, 60 Garden Street MS-6, Cambridge, MA 02138;
bgaensler@cfa.harvard.edu}
\altaffiltext{2}{Harvard College, Cambridge MA 02138}
\altaffiltext{3}{School of Physics and Astronomy, University of St Andrews,
St Andrews, Fife KY16 9SS, United Kingdom}
\altaffiltext{4}{Department of Astronomy, University of Florida, 
211 Bryant Space Science Center, Gainesville FL 32611}
\altaffiltext{5}{Physics Department and Center for Space Research,
Massachusetts Institute of Technology, 70 Vassar Street, Cambridge MA 02139}
\hspace{-5cm}

\begin{abstract}

We report on the detection of X-ray emission from the unusual Galactic
radio source \tornado\ (the ``Tornado'').  Observations made with the
{\em Chandra X-ray Observatory}\ demonstrate the presence of up to three
sources of X-ray emission from the Tornado: a relatively bright region
of dimensions $2'\times1'$ coincident with and interior to the brightest
radio emission at the ``head'' of the Tornado, plus two fainter extended
regions possibly associated with the Tornado's ``tail''.  No X-ray point sources
associated with the Tornado are seen down to a $3\sigma$ luminosity
(0.5--10~keV) of $1\times10^{33}$~erg~s$^{-1}$, for a distance to the
system of 12~kpc.  The spectrum of the brightest region of X-rays is
consistent with a heavily absorbed ($N_H \approx 10^{23}$~cm$^{-2}$)
thermal plasma of temperature $kT \sim 0.6$~keV; an absorbed power law
can also fit the data, but implies an extremely steep photon index. From
these data we tentatively conclude that the Tornado is a supernova
remnant (SNR), although we are unable to rule out the possibility that
the Tornado is powered either by outflows from an X-ray binary or by the
relativistic wind of an unseen pulsar. Within the SNR interpretation,
the head of the Tornado is a limb-brightened radio shell containing
centrally-filled thermal X-rays and which is interacting with a molecular
cloud.  We therefore propose that the Tornado is a ``mixed morphology''
supernova remnant.  The unusual tail component of the Tornado remains
unexplained in this interpretation, but might result from expansion of
the SNR into an elongated progenitor wind bubble.

\end{abstract}


\section{Introduction}
\label{sec_intro}

The bright radio source \tornado\ is extended, has high levels of
polarization and shows a non-thermal spectrum, properties which at
first led to its classification as a supernova remnant (SNR). However,
high-resolution images revealed a bizarre axial ``head-tail'' structure
(\cite{ssp+85}; \cite{hb85}; top panel of Fig~\ref{fig_tornado}), for
which this source was nicknamed ``the Tornado''. The morphology of
the Tornado is very different from the approximately circular morphologies
seen in other SNRs. This has led to various  claims that the Tornado is
an exotic SNR expanding into an unusual environment, is a nebula powered
by a high velocity pulsar, or results from precessing jets from an X-ray
binary (e.g.\ \cite{man87}; \cite{sfs89}; \cite{shgr94}).

Recent observations have provided some new clues.
Frail \etal\ (1996\nocite{fgr+96}) detected a 1720-MHz OH maser
from the ``head'' of the Tornado.  The 1720~MHz emission, when seen
without the other OH maser lines, is produced by 
shocks propagating into molecular clouds; further evidence for
an interaction with molecular material is provided by the presence of
emission from $^{13}$CO 1--0 and shock-excited 2.12-$\mu$m H$_2$ at
this same position and velocity (\cite{lby+03}).  The maser velocity,
also now confirmed by \HI\ absorption measurements (\cite{bg03}), implies
a distance to the Tornado of 12~kpc.

Despite these new data, the Tornado still defies
classification. Most problematic has been the failure to detect
emission in other wavebands.  We consequently
examined archival {\em ASCA}\ data of this region, and identified a weak
($\sim3\sigma$) X-ray source coincident with the Tornado's head;
this marginal detection was also briefly noted by Yusef-Zadeh \etal\
(2003\nocite{ywrs03}). We here present an observation of the Tornado with
the {\em Chandra X-ray Observatory}, aimed at confirming
this source of X-rays, and studying its morphology and spectrum.

\section{Observations and Results}

The Tornado was observed with \cxo\ on 2002 May 07. We used the ACIS-I
detector, consisting of an array of four front-illuminated CCDs and giving
a field of view approximately $17'\times17'$. The aimpoint, located
near the center of the array, corresponded to coordinates RA (J2000)
$17^{\rm h}40^{\rm m}15\fs33$, Decl.\ (J2000) $-30^\circ57'59\farcs2$.
The effective exposure time was 20\,723~seconds, and was free
of background flares.

\subsection{Imaging}
\label{sec_image}

\begin{figure}
\centerline{\psfig{file=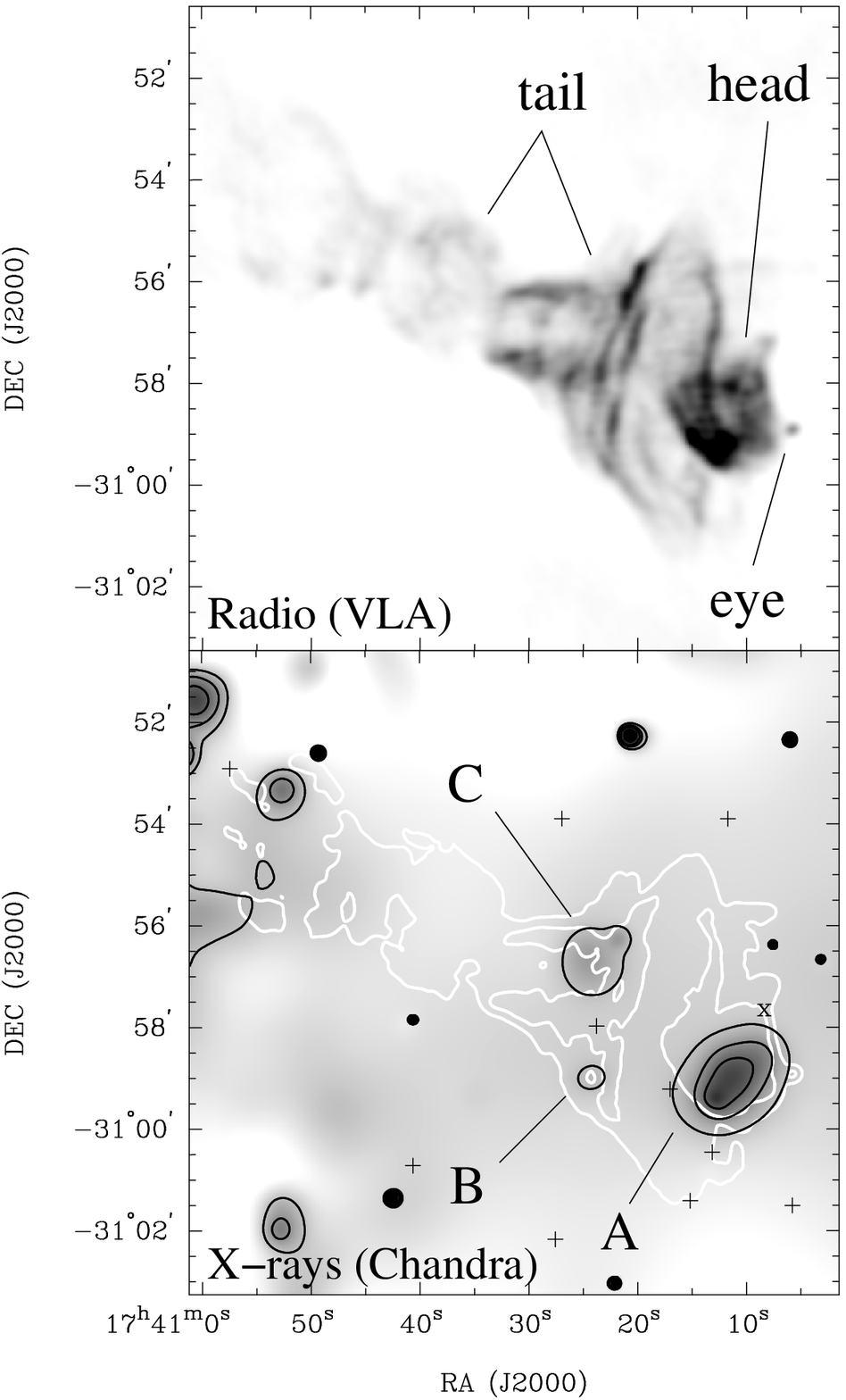,height=0.75\textheight}}
\caption{Radio and X-ray images of \tornado\ (``the Tornado''). 
The top panel shows a
1.4-GHz VLA image of the Tornado, corrected for primary beam attenuation
and with a spatial resolution of $14''\times11''$ (\cite{bg03}).
(The ``eye'' of the Tornado, is an unrelated \HII\ region; \cite{bg03}.)
The image in the bottom panel 
shows \cxo\ ACIS-I data in the energy range 0.3--10~keV
after exposure correction and adaptive smoothing.  In the bottom panel,
black contours represent X-ray emission at levels of 76\%, 84\% and 92\%
of the peak brightness of region~A, while
white contours represent radio emission at levels of 7.5 and
50~mJy~beam$^{-1}$.
Point sources identified with {\tt wavdetect}\ but
which are too faint to see in the smoothed image are marked with
``+'' symbols. The position of the OH maser
is marked with a ``X'' symbol.
X-ray emission seen in the northeast and southeast
corners of the image is due to the large exposure correction near the
edges of the field of view, and does not correspond to real emission.
Regions discussed in the text are indicated.}
\label{fig_tornado}
\end{figure}

Data were analyzed using {\tt CIAO}\ v2.3.  We used the level-1
events supplied by the Chandra X-ray Center (CXC), and then corrected
them for charge-transfer inefficiency using standard CXC routines.
An image was then formed using events in the energy range 0.3 to 10~keV;
point sources in this image were identified using the {\tt wavdetect}\
algorithm within {\tt CIAO}. To  highlight diffuse emission, we applied
the adaptive smoothing algorithm {\tt csmooth}.  We corrected for
off-axis vignetting and for the variable exposure produced by chip gaps
and bad pixels by forming an exposure map, which we then
smoothed  with the same kernel as was applied to the
image. Applying the smoothed exposure map to the smoothed image resulted
in a flux-calibrated image of diffuse emission.  The resulting X-ray image
is shown in the lower panel of Figure~\ref{fig_tornado}, demonstrating
that faint emission is detected from three distinct regions coincident
with the Tornado, which we label A, B and C.

The brightest X-rays are from region A, lying near the ``head''
of the Tornado, and comprising a $\sim2'\times1'$ elliptical region
elongated along a position angle $\sim150^\circ$ (north
through east).  Comparison with radio data shows that these X-rays are
wholly contained within the head region, occupying the western half of
the head's interior.  While region A is brighter and more
concentrated at its south-eastern tip, no unresolved source is identified
in this region by {\tt wavdetect}, nor is apparent by visual inspection
of the unsmoothed image.

Regions B and C represent two other possible sources of extended X-rays
from the Tornado, located $\sim2'$ east of the head region.  Region B is
a very faint circular clump of extent $\sim30''$, while region C 
is a brighter and more elongated feature of extent $\sim1'$. Given
the faintness of regions B and C, we simply note their existence, and
defer detailed studies of these regions until deeper observations have
been performed.
The {\tt wavdetect}\ algorithm identifies two X-ray point sources
lying within the radio boundaries of the Tornado, as marked in
Fig~\ref{fig_tornado}. Both sources are of marginal significance, each
containing only 5--6 counts.

\subsection{Spectroscopy}

Of the three extended regions of X-ray emission identified in
\S\ref{sec_image},
only region A is bright enough from which to obtain a useful spectrum.
Events  were extracted from an elliptical extraction region of dimensions
$150'' \times 120''$ centered on this emission, and  the data 
(source plus background) were then
regrouped so that there were at least 50 counts in each energy bin.
The effective area for this source spectrum was determined by weighting
by the brightness of the source over the extraction region
using the {\tt CIAO} task {\tt mkwarf}.  The local
background was determined by extracting a spectrum from an adjacent
emission-free region of the CCD immediately to the southwest of region~A.

The corresponding spectrum contains $162\pm37$ background-subtracted
counts, which fall into 12 bins spread over the energy range
0.3--10~keV. We fit two simple models to this spectrum: a power law, and
a Raymond-Smith plasma in collisionally ionized equilibrium,\footnote{More
complicated thermal models including non-equilibrium ionization give near
identical best-fit parameters.} both modified by foreground photoelectric
absorption.  Both models give good fits to the data, the parameters for
which are listed in Table~\ref{tab_spec}. However, the implied photon
index for the best power law fit is unusually steep, $\Gamma > 4.5$
at 90\% confidence.  In Table~\ref{tab_spec} we also present power
law fits for fixed photon indices $\Gamma = 2.3$ and $\Gamma = 3.5$
(representing extreme value for observed sources; see \S\S\ref{sec_pwn}
\& \ref{sec_snr}), which give comparatively poor fits to the data.

\section{Discussion}

We have clearly identified X-ray emission from the head of the Tornado;
the observed count-rate is consistent (to within $\sim50$\%) with the
level of emission suggested from archival {\em ASCA}\ data.  Although we
detect relatively few X-ray photons from the Tornado, these data provide
new constraints on the nature of this complicated source.

\subsection{Outflows from an X-ray binary?}
\label{sec_binary}

Stewart \etal\ (1994\nocite{shgr94}) propose that the Tornado is powered
by twin outflows from a central X-ray binary, presumably
located in the center of and along the symmetry axis of the overall
radio nebula, and which generate opposed precessing jets as is
seen for SS~433 in the SNR~W50 (e.g.\ \cite{mar84b}).  We argue
in \S\ref{sec_snr} below that region~A has a thermal spectrum.  In the
context of the situation proposed by Stewart\nocite{shgr94} et al.\ (1994),
these thermal X-rays might represent the working surface where a
collimated outflow from the central source collides and interacts with
surrounding gas, as is seen at the termination of the eastern lobe of
W50 (\cite{bak96}; \cite{so97}).

Of immediate concern for such a model is that no bright unresolved
X-ray emission, corresponding to a central powering source, is seen
along the axis of the Tornado.  The $3\sigma$ upper limit on
the count-rate from any such source is $\approx5\times10^{-4}$~cts~s$^{-1}$
in the energy range 0.3--10~keV. Assuming a foreground absorbing column
density of $N_H \approx 1 \times 10^{23}$~cm$^{-2}$ (see \S\ref{sec_snr} below)
and a power-law spectrum with photon index $\Gamma = 2$, we infer an
unabsorbed flux (0.5--10~keV) for such a source $f_X \la 6 \times
10^{-14}$~ergs~cm$^{-2}$~s$^{-1}$, which implies an upper limit on the
X-ray luminosity (0.5--10~keV) $L_X \la 1\times10^{33}$~ergs~s$^{-1}$
at a distance of 12~kpc.

From this non-detection an exact analogy with SS~433 can immediately
be ruled out: even when not in outburst, such a source would have an
X-ray luminosity $L_X \sim 10^{36}$~ergs~s$^{-1}$ and a 1.4-GHz flux
density $\sim100$~mJy (see \cite{mar84b} and references therein), 
which would have been easily detected in \cxo\ and VLA observations
of the Tornado, respectively.  However, we below note two other X-ray
binaries known to generate outflows, which if embedded in the Tornado
might not be directly detectable.

The recently discovered high-mass X-ray binary LS~5039 generates
steady jets of velocity $\sim0.2c$ and kinetic luminosity
$\sim10^{37}$~ergs~s$^{-1}$ (\cite{pmrm00}), which is consistent with the
rate of energy input required to power the radio emission seen from the
Tornado (\cite{hb85}).  The X-ray luminosity of LS~5039 possibly varies as
a function of orbital phase, but at its minimum is only slightly higher
than the upper limit determined for any point source in the Tornado
(\cite{rrpm03}).  At a distance of 12~kpc, the 1.4-GHz flux density of
this source would only be 1--3~mJy (\cite{mpr98}), too faint to be seen
in existing VLA radio images of this region.

The low-mass X-ray binary 4U~1755-338 is currently in a quiescent state
with $L_X < 4 \times 10^{31}$~~ergs~s$^{-1}$, but 
previously underwent an active phase lasting several decades,
in which it generated X-ray jets $\sim$4~pc in extent (\cite{aw03}).
The Tornado could similarly have been powered by a period of
prolonged activity, but is now quiescent. Jets of the surface brightness
seen for 4U~1755-338 would be too faint to be detected here.


We note that an alternative X-ray binary model 
has been proposed by Helfand \& Becker (1985\nocite{hb85}),
who argue that the Tornado is powered by source at the very
western edge of the head, and that the high velocity of the binary then
gives the Tornado its unusual appearance. The arguments invoked
above can explain the failure to detect X-ray emission from
the  binary itself.  However, in this model the energetic interaction
between the outflow and its surroundings should occur only at the very
western edge of the Tornado.  In such a situation, there is no simple
explanation for the presence of extended X-ray emission from region A,
since this should be neither a region of active particle acceleration
nor an area in which shock-heated gas is produced. We therefore think
this possibility unlikely.

\subsection{A pulsar powered nebula?}
\label{sec_pwn}

Shull \etal\ (1989\nocite{sfs89}) have argued that the Tornado is a pulsar
wind nebula (PWN) powered by a high-velocity pulsar. In such a situation,
the pulsar is presumed to be inside of or to the west of the head of the
Tornado. We can consequently expect to see extended synchrotron X-ray
emission from a surrounding PWN, plus possibly unresolved magnetospheric X-ray
emission from the pulsar itself.

Region A clearly corresponds to X-ray emission in the expected location
for this interpretation, and its spectrum can be well-fitted by a
power law characteristic of synchrotron emission. However, all known
PWNe show power law X-ray spectra with photon indices in the range
$1.3 \la \Gamma \la 2.3$ (\cite{got03}); Table~\ref{tab_spec} shows
that a photon index even at the upper limit of this range, $\Gamma =
2.3$, gives a comparatively poor fit, which we can exclude at the
$\sim2.3\sigma$ level.  While more data are needed to confirm whether
this source indeed has an anomalously steep spectrum, this measurement
suggests that the properties of region A are not consistent with those
expected for a pulsar-powered nebula.  We note that while we detect
no point source embedded in region~A, the corresponding upper limit
$L_X \la 1\times10^{33}$~ergs~s$^{-1}$ (see \S\ref{sec_binary} above)
is above the X-ray luminosities of many young pulsars (\cite{pccm02}).

\subsection{A supernova remnant}
\label{sec_snr}

The Tornado shows bright, polarized, limb-brightened radio emission
with a non-thermal spectrum, properties which resulted in its original
classification as a SNR. However, one must then explain its  bizarre
and unique morphology. 

In a dense ambient medium a supernova progenitor's space velocity can
take it near the edge of, or even outside of, its circumstellar bubble
before the star explodes. The SNR resulting from such a system will
have a highly elongated axial morphology (\cite{rtfb93}; \cite{bd94}),
resembling that seen for the Tornado.  The OH maser identified towards
this source by Frail \etal\ (1996\nocite{fgr+96}), along with the recent
detection of shock-excited 2.12-$\mu$m H$_2$ emission and surrounding
$^{13}$CO emission by Lazendic \etal\ (2003\nocite{lby+03}), all argue
that indeed the Tornado is expanding into a dense environment.

For a SNR interpretation there are several different possibilities
for the X-ray emission expected. We might see limb-brightened
synchrotron emission from shock-accelerated electrons (e.g.\ 
G347.3--0.5; \cite{sgd+99}), limb-brightened thermal emission
from shock-heated ambient gas (e.g.\ 1E~0102.2--7219; \cite{hrd00}),
or centrally-filled thermal emission from hot gas in the interior,
corresponding to a so-called ``mixed morphology'' or ''thermal composite'' 
SNR (e.g.\ W44; \cite{rpsh94}). 

The X-ray photon index for synchrotron-emitting SNRs is in the range $2.5
\la \Gamma \la 3.5$ (see \cite{pet01b} for a review)
which, as for the case of a PWN in \S\ref{sec_pwn}, is distinctly
flatter than the best-fit power law observed. On the other hand,
the temperature of $kT \sim 0.6$~keV implied by thermal
models is quite reasonable for thermally emitting SNRs, as is the
corresponding unabsorbed X-ray luminosity (0.5--10~keV) of $L_X \sim 
10^{36}$~erg~s$^{-1}$ for a distance of 12~kpc.  The foreground
column $N_H \approx 10^{23}$~cm$^{-2}$ is high, but is well
below the total integrated column in this direction of $N_H
\approx 1.5\times10^{23}$~cm$^{-2}$ (\cite{dl90}).

If we identify the head of the Tornado as a limb-brightened radio SNR,
then the emission seen with \cxo\ originates wholly from the SNR
interior, and we correspondingly can tentatively classify the Tornado as a
mixed morphology SNR (as had been earlier speculated by Yusef-Zadeh \etal\
2003\nocite{ywrs03} from the possible presence of {\em ASCA}\ emission).
Whether such centrally-filled thermal X-rays are produced by evaporation
of dense clouds overrun by the shock, through thermal conduction in
the interior, or via some other process is currently
controversial (\cite{rp98}).  However, regardless of the mechanism invoked,
there have been repeated arguments that such a SNR is the result of an
interaction with a dense molecular cloud (\cite{gfgo97}; \cite{rp98};
\cite{ywrs03}).  The OH maser, 2.12-$\mu$m H$_2$ and $^{13}$CO emission
seen here all support this interpretation.

Wardle (1999\nocite{war99}) has pointed out that the soft X-ray emission
produced by a mixed morphology SNR can produce the ionizing flux $\zeta
\ga 10^{-16}$~s$^{-1}$ needed to disassociate H$_2$O and
produce the OH seen in maser emission. At a distance of 12~kpc,
the maser produced by the Tornado is $\sim3.5$~pc
from the source of X-rays, implying a flux as seen
by the OH maser $f_X \sim 6\times10^{-4}$~ergs~cm$^{-2}$~s$^{-1}$.
The consequent ionizing flux is $\zeta = N_e \sigma f_X \sim
3\times10^{-15}$~s$^{-1}$, where $N_e \approx 30$~keV$^{-1}$ is
the number of electrons produced per keV of ionization and
$\sigma \approx 2.6\times10^{-22}$ is the photoabsorption
cross-section of hydrogen at 1~keV (\cite{war99}). This
rate of ionization
is more than sufficient to convert water into OH molecules.

\section{Conclusions}

A short observation of \tornado\ with the {\em Chandra X-ray Observatory}\
has confirmed the presence of X-ray emission from this remarkable object.
The brightest region of X-rays from this source is clearly
extended, seems to sit within the shell of emission defined by the
``head'' of the radio source, and has a highly absorbed spectrum
consistent with being either an extremely steep power law or a
collisionally ionized plasma at a temperature $\sim0.6$~keV.

We have considered the possibility of the Tornado being powered by
jets from an X-ray binary, of corresponding to the synchrotron nebula
powered by an energetic rotation-powered pulsar, or of representing
an unusually-shaped supernova remnant.  While the available data
do not allow us to definitively rule out any of these alternatives,
the spectrum and morphology of the X-ray emission seen by \cxo, when
combined with evidence at other wavebands for the interaction of a shock
with a molecular cloud, all suggest that the head of the Tornado is a
``mixed morphology'' supernova remnant, producing thermal X-rays from
the interior of a limb-brightened radio shell, and expanding into a
dense and possibly complicated distribution of ambient gas.

This argument provides a reasonably simple interpretation for the
``head'' of the Tornado, but the bizarre morphology of the ``tail''
remains unexplained. We speculate that the Tornado might result from
expansion of a SNR into a highly elongated progenitor wind bubble, but
explanations involving pre-existing helical magnetic fields, or high-speed
outflows from a still unseen central source, can also not be discounted.
Deep X-ray observations with \cxo\ and {\em XMM}\ are needed to
better determine the morphology and spectrum of X-rays from the head,
and to study the possible X-ray emission from the tail.


\begin{acknowledgements}

We thank Dan Harris, Jasmina Lazendic and Fred Seward for useful
discussions, and Crystal Brogan for supplying us with her radio images of
the Tornado.  This work was supported by NASA through SAO grant GO2-3041X
(BMG) and by the NSF (JMM).

\end{acknowledgements}



\begin{table}[htb]
\begin{center}
\caption{Spectral fits to X-ray emission from the head of the Tornado.}
\label{tab_spec}
\begin{tabular}{ccccc} \hline 
Model & $N_H$ ($10^{22}$~cm$^{-2}$) & $\Gamma$ / $kT$ (keV) &
$f_x$ ($10^{-11}$~erg~cm$^{-2}$~s$^{-1}$) &
$\chi_\nu^2/\nu$ \\ \hline
RS & $10^{+7}_{-4}$ & $0.6^{+0.8}_{-0.3}$ & $\sim5.1$ &
$10.2/9 = 1.1$ \\ 
PL &  $12^{+6}_{-5}$ & $>4.5$ & $>1.2$ & $10.8/9=1.2$ \\
PL &  $\sim3.4$ & $2.3$ (fixed)\tablenotemark{a}  & $\sim0.05$ & $20.3/10 = 2.0$ \\ 
PL &  $5\pm1$ & $3.5$ (fixed)\tablenotemark{a}  & $\sim0.2$ & $15.7/10 = 1.6$ \\ \hline
\end{tabular}
\tablenotetext{a}{See \S\S\ref{sec_pwn} and \ref{sec_snr} for details.}
\tablenotetext{}{Uncertainties and lower limits are all at 90\% confidence.
All models assume interstellar absorption using the
cross-sections of Ba\protect\mbox{\l}uci\protect\'{n}ska-Church 
\& McCammon (1992\protect\nocite{bm92}), assuming solar abundances.
Models used: ``PL'' indicates a power law of the
form $f_\varepsilon \propto \varepsilon^{-\Gamma}$ where $\Gamma$ is the
photon index;
``RS'' indicates a Raymond-Smith spectrum of
temperature $T$ (\protect\cite{rs77}).
Fluxes are for the energy range 0.5--10~keV,
and have been corrected for interstellar absorption.}
\end{center}
\end{table}

\normalsize

\clearpage

\end{document}